\documentclass[prb,twocolumn,superscriptaddress]{revtex4-1}
\usepackage{graphicx,amsmath,amssymb,bm,placeins}
\usepackage{hyperref}
\newcommand{\bsigma}{{\boldsymbol\sigma}}
\newcommand{\bk}{{\bm{k}}}
\newcommand{\bg}{{\bf g}}
\newcommand{\tp}{{t_{\perp}}}
\newcommand{\dkx}{{\partial_{k_x}}}
\newcommand{\dky}{{\partial_{k_y}}}
\newcommand{\imag}{{\mathrm{Im}}}

\begin{document}
\title{Marginal topological properties of graphene: a comparison with
topological insulators}
\author{Jian Li}
\affiliation{D\'{e}partement de Physique Th\'{e}orique,
Universit\'{e} de Gen\`{e}ve, CH-1211 Gen\`{e}ve 4,
Switzerland}
\author{Ivar Martin}
\affiliation{Theoretical Division, Los Alamos National Laboratory, Los Alamos,
New Mexico 87545, USA}
\author{Markus B\"{u}ttiker}
\affiliation{D\'{e}partement de Physique Th\'{e}orique,
Universit\'{e} de Gen\`{e}ve, CH-1211 Gen\`{e}ve 4,
Switzerland}
\author{Alberto F. Morpurgo}
\affiliation{DPMC and GAP, Universit\'{e} de Gen\`{e}ve, CH-1211 Gen\`{e}ve 4,
Switzerland}

\begin{abstract}
The electronic structures of graphene systems and
topological insulators have closely-related features, such
as quantized Berry phase and zero-energy edge states.
The reason for these analogies is that in both systems there are two relevant orbital bands, which generate the pseudo-spin degree of freedom, and, less obviously, there is a correspondence between the valley degree of freedom in graphene and electron spin in topological insulators.
Despite the similarities, there are also several important distinctions, both for the
bulk topological properties and for their implications for the
edge states -- primarily due to the fundamental difference between
valley and spin. In view of their peculiar band structure features, gapped graphene systems should be properly characterized as marginal topological insulators, distinct from either the trivial insulators or the true
topological insulators.
\end{abstract}
\maketitle

\section{Introduction}

Condensed matter physics is witnessing an increasingly rapid development marked with a sequence of surprising discoveries. Two good recent
examples are the realization of graphene \cite{geim_rise_2007} -- a plane of carbon atoms forming honeycomb lattice, and topological
insulators \cite{hasan_topological_2010, qi_quantum_2010} -- materials that are insulating in the bulk but conducting at the surface owing
to topological reasons. In fact, the two examples were linked since the beginning: quickly after the initial experiments on graphene
\cite{novoselov_two-dimensional_2005, zhang_experimental_2005}, it was suggested by Kane and Mele, that graphene may be gapped due to the
intrinsic spin-orbit interaction and provide a prototype of a novel class of time-reversal-invariant topological insulators
\cite{haldane_model_1988, kane_quantum_2005, kane_z2_2005}.

This idea has since inspired numerous works that have surprisingly, yet significantly, complemented our knowledge of some seemingly well
developped fields of solid state physics. The original prediction by Kane and Mele, however, remains unfulfilled in graphene because of the
very weak spin-orbit interaction in realistic graphene samples.

In fact, in graphene, spin-orbit interaction is so weak that
in most existing experiments without a magnetic field, the
spin degree of freedom hardly plays a discernible role
except for contributing an additional degeneracy. This
certainly prevents graphene from being an ideal
representative of (time-reversal-invariant) topological
insulators for which strong spin-orbit interaction is
normally a crucial ingredient. Despite this, graphene and
topological insulators share a number of closely related
features like a quantized Berry phase
\cite{novoselov_two-dimensional_2005,
zhang_experimental_2005, novoselov_unconventional_2006} and
zero-energy edge states \cite{castro_localized_2008,
martin_topological_2008, semenoff_domain_2008,
yao_edge_2009}. Both in graphene and topological insulators
one can isolate two relevant bands with the opposite orbital
symmetry, which in the band structure are connected by the
matrix elements linear in momentum, leading to the Dirac
cones.  Further, in both systems, a gap in the Dirac
dispersion can be opened, by means of spin-orbit interaction
in topological insulators, and simply by creating a
sublattice on-site energy imbalance in graphene. The analogy
between the two systems is completed by associating electron
spin in topological insulators with the valley degree of
freedom (also time-reversal odd) in graphene.
 At the formal level, as we shall see in a moment, the effective theories for valleys and for spins
bear such great similarities, that one could very well be tempted to transplant the conclusions obtained in topological insulators into
graphene systems, with spin substituted by valley. This raises the question: how far exactly can such a spin and valley analogy be pushed?

To examine carefully the analogies and the differences between valley-based graphene-like systems and spin-based topological insulators, it is
helpful to compare specific examples which nevertheless possess generic properties. For the graphene case we will use
electrostatically-biased bilayer graphene (BLG) \cite{castro_biased_2007}, where an electronic bulk band gap can be opened in a practical
and controllable manner \cite{oostinga_gate-induced_2008, zhang_direct_2009, mak_observation_2009}. For the topological insulator case, we
will use the Bernevig, Hughes and Zhang (BHZ) model \cite{bernevig_quantum_2006}, which is a good prototype of two-dimensional (2D)
topological insulators and has its realization in HgTe quantum wells \cite{konig_quantum_2007, koenig_quantum_2008, roth_nonlocal_2009}. Our
comparison will be focused on the characterization of bulk properties and their relation to the presence of subgap edge modes. This scheme
follows intimately the fundamental logic underlying the idea of topological insulators, namely, the bulk-edge correspondence.
\cite{hatsugai_chern_1993, qi_general_2006} The outcome of this investigation has direct experimental implications in terms of subgap edge
transport.

We emphasize that the purpose of this paper is by no means to provide a stringent and comprehensive review on the comparison between
graphene systems and topological insulators -- as one of the limitations, we will ignore in the following the spin degree of freedom in
graphene. We highlight several points that illustrate the analogies and the differences behind valley numbers and spin. The manner of our
presentation is intended to be heuristic. We refer the readers to Ref. \cite{li_marginality_2010, li_topological_2011} for more details and
more rigor.

\section{Analogy between graphene systems and topological insulators}\label{sec:analogy}

In this section we demonstrate the analogy between graphene systems and 2D topological insulators by comparing two examples: gapped BLG and
the BHZ topological insulator. In both examples, we consider for the moment the ideal cases -- meaning, for gapped BLG we ignore coupling
between valleys, and for the BHZ topological insulator we assume spin is a good quantum number.

\subsection{Effective Hamiltonians}\label{ssec:hams}

We start by writing down the low-energy effective Hamiltonian for each system. For gapped BLG, this reads
\begin{align}
&H_{BLG}(\bk) = \left(%
\begin{array}{cc}
  H_K(\bk) & 0 \\
  0 & H_{K'}(\bk) \\
\end{array}%
\right), \label{eq:ham_blg2v} \\ &H_K(\bk) = -\left(%
\begin{array}{cc}
  \Delta & (k_x - ik_y)^2 \\
  (k_x + ik_y)^2 & -\Delta \\
\end{array}%
\right), \label{eq:ham_blgK} \\ &H_{K'}(\bk) = H_K^{\ast}(-\bk), \label{eq:ham_blgKp}
\end{align}
where $K$ and $K'$ denote the two valleys, $\bk$ is the wave vector relative to each valley, and $\Delta$ is the electrostatically tunable
gap; for the BHZ topological insulator, it reads (with immaterial simplifications of the original model)
\begin{align}
&H_{BHZ}(\bk) = \left(%
\begin{array}{cc}
  H_S(\bk) & 0 \\
  0 & H_{S'}(\bk) \\
\end{array}%
\right), \label{eq:ham_bhz2s} \\ &H_S(\bk) = \left(%
\begin{array}{cc}
  m-k^2 & a(k_x - ik_y) \\
  a(k_x + ik_y) & -(m-k^2) \\
\end{array}%
\right), \label{eq:ham_bhzS} \\ &H_{S'}(\bk) = H_S^{\ast}(-\bk), \label{eq:ham_bhzSp}
\end{align}
where $S$ and $S'$ denote the two spin sectors, $m>0$ and $a\ne0$ are two real parameters. Note that, for simplicity, both Hamiltonians
above have been made dimensionless by choosing fixed length and energy scales in each model.

One immediate observation can be made regarding the similar
roles played by valley and spin. In the BHZ topological
insulator, time reversal symmetry is respected and imposes a
constraint on the two spin sectors given by Eq.
\eqref{eq:ham_bhzSp}. The same constraint, given by Eq.
\eqref{eq:ham_blgKp}, governs the two valley sectors of
gapped BLG also as a consequence of time reversal invariance
(as mentioned above, due to lack of any significant
spin-orbit interaction spin degree of freedom is decoupled
from the orbital ones; therefore one can treat electrons as
spinless particles in all calculations, reinstating the spin
at the end). These constraints, generally obeyed in the two
systems \cite{note1}, have crucial implications for
the overall bulk and edge properties as we will see in due
course.

\begin{figure}
\begin{center}
  \includegraphics[width=0.35\textwidth,height=10cm]{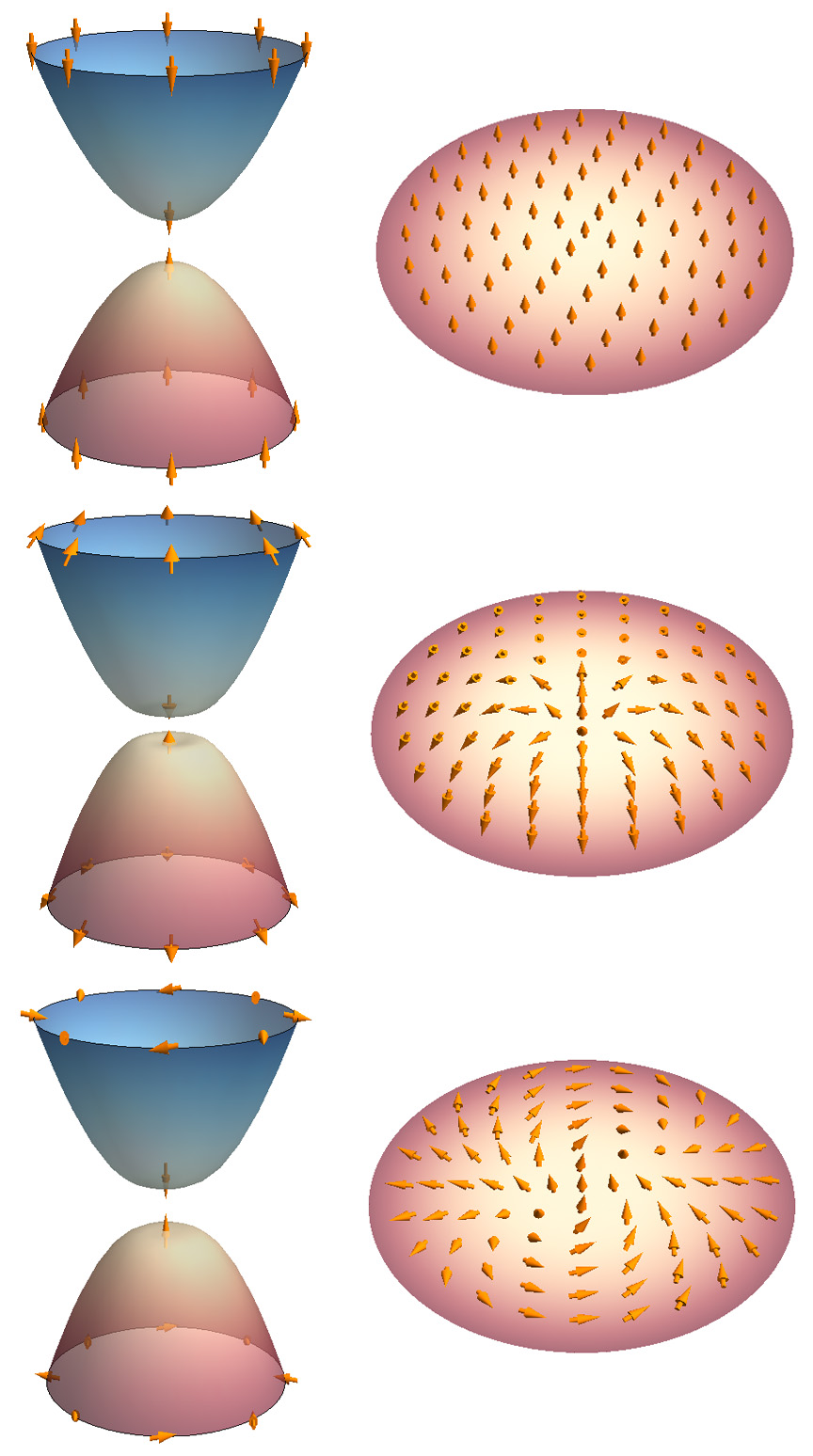}
\end{center}
\caption{The (pseudo-)spin textures for the eigenstates of:
a trivial two-band system (top panels), one spin sector of
the BHZ topological insulator (cental panels), and one
valley of gapped BLG (bottom panels). The left panels show
the band structures at low energy and the spin polarization
for some selected states; the right panels show more details
of the lower bands. Note that, for illustration purpose, the
Hamiltonians $H(\bk)=-\bg(\bk)\cdot\bsigma$ (with an extra
minus sign as compared with those in the text) are used such
that the spin polarization of the lower band states is
aligned with $\hat{\bg}$. The trivial system is defined by
setting $\bg(\bk)=(0,0,m+k^2)$ with $m>0$.}
\label{fig:spin_all}
\end{figure}

The two effective Hamiltonians, when characterized using the usual topological measures, seem to show an even closer analogy. For this
purpose we use the common expression for the topological invariant associated with a fully-gapped two-component Hamiltonian
$H(\bk)=\bg(\bk)\cdot {\bsigma}$. Here $\bsigma$ is the vector of the Pauli matrices and $\bg(\bk)$ is a real vector. In terms of the
normalized vector $\hat{\bg}= {\bg}/|{\bg}|$ the expression for the topological invariant is \cite{volovik_universe_2003}
\begin{align}\label{eq:cnum}
c = \frac{1}{4\pi}\int {d^2k} \:\hat{\bg} \cdot \left( \dkx \hat{\bg} \times \dky \hat{\bg} \right)\,.
\end{align}
The number $c$ characterizes the mapping from the 2D $k$-space to the 2D parametric sphere subtended by $\hat{\bg}$ (see Fig.
\ref{fig:spin_all}; for details see Section \ref{sec:lim}). It is straightforward to find for individual valley/spin sectors (assuming
$\Delta<0$; $c_{K/K'}$ changes sign when $\Delta$ changes sign)
\begin{align}
&c_K = -c_{K'} = 1, \label{eq:cnum_blg} \\ &c_S = -c_{S'} = 1 \label{eq:cnum_bhz}.
\end{align}
Namely, the signs of $c$'s reverse with respect to valleys/spins, echoing the underlying symmetry; the magnitudes of $c$'s appear to be the
same integer $1$ in both gapped BLG and the BHZ topological insulator!

\subsection{Band inversion}\label{ssec:band}

The nontrivial bulk properties characterized by Eqs. \eqref{eq:cnum_blg} and \eqref{eq:cnum_bhz} are closely related to the occurrence of
band inversion. Here we present a simple account of the band inversion in each system, bearing in mind that such accounts are by no means
descriptions of the microscopic mechanisms, but rather phenomenological illustrations based on the effective models.

\begin{figure}
\begin{center}
  \includegraphics[width=0.42\textwidth]{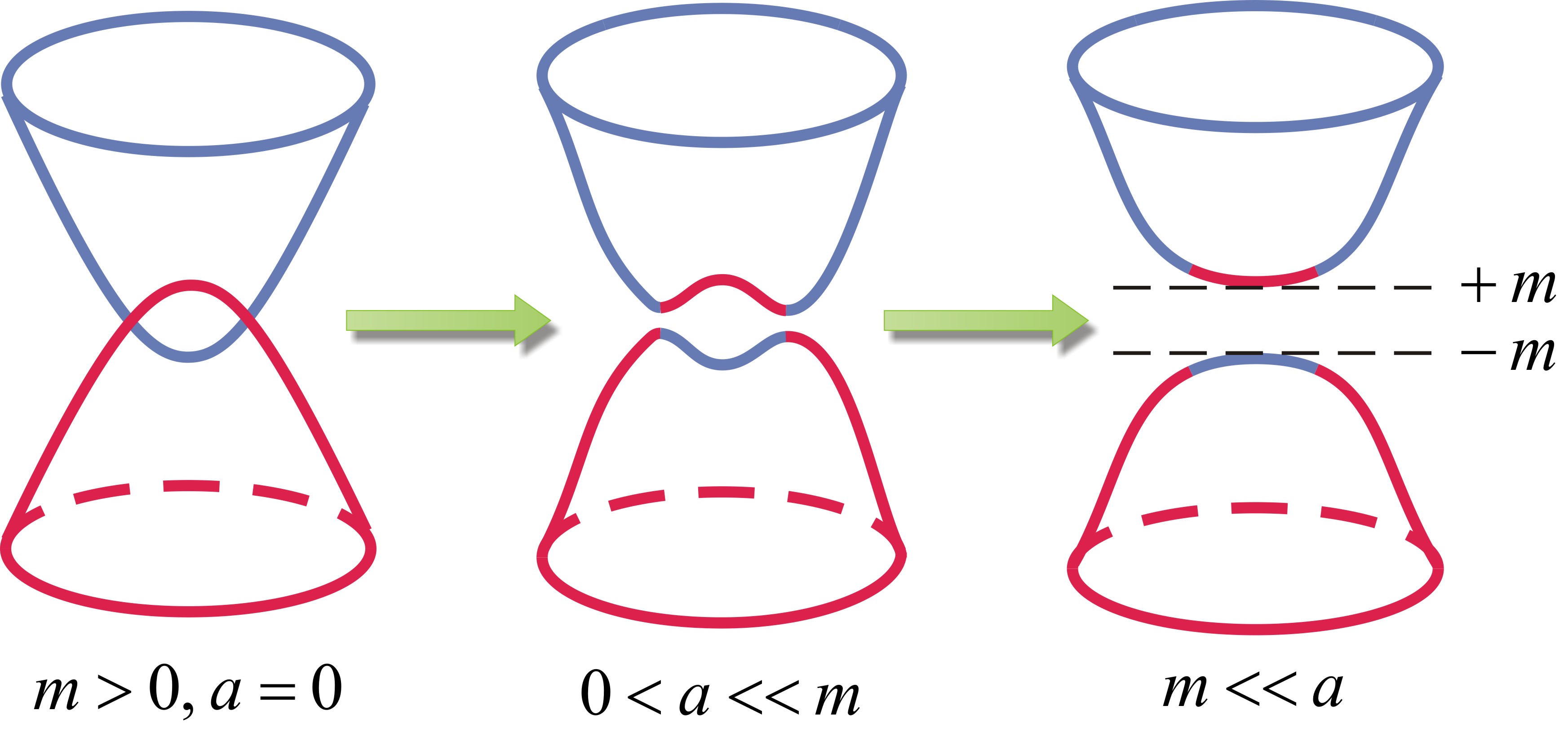}
\end{center}
\caption{Evolution of the band structure corresponding to the Hamiltonian \eqref{eq:ham_bhzS} by tuning $a$, with fixed positive $m$. This
illustration is to be compared with Fig. \ref{fig:inv_blg}  to provide a naive analogy between the gap opening in BLG and that in the BHZ
insulator (see text for details). Note that the comparison between $a$ and $m$ should be understood as being under a certain fixed length
scale which converts the two parameters to the same dimension.} \label{fig:inv_bhz}
\end{figure}

We start with the BHZ topological insulator. Due to the symmetry, it is sufficient to look at one spin sector, $H_S$ say. Let us first
compare two cases, with $m<0$ and $m>0$ respectively, and $a=0$. It is obvious that the eigenstates in both cases are completely polarized
in terms of the two-component pseudo-spin. Meanwhile the crucial difference is that in the latter case ($m>0$), the two parabolic bands
intersect each other, while in the former case ($m<0$) they are fully separated. In a purely phenomenological sense, we call such a reversal
of the order of bands at low energy ``band inversion". But the real excitement about band inversion comes only after we turn on $a$, which
leads to an avoided crossing between the two bands resulting in the opening of a gap (which makes the bulk of the system insulating). Fig.
\ref{fig:inv_bhz} shows how the profile of the inverted bands changes with increasing magnitude of $a$. Clearly enough, at small $k$ (where
only $m$ plays an important role), the electronic eigenstates tend to preserve their inverted character -- states in the upper/lower band
are largely polarized (in terms of pseudo-spin, see also Fig. \ref{fig:spin_all}) the same way as the original downward/upward parabolic
band; at large $k$ (where both $m$ and $a$ become irrelevant), the order restores and the eigenstates in the upper/lower band deviate little
from the original upward/downward parabolic bands; in-between, as long as $a\ne0$, the eigenstates continuously evolve such that the
pseudo-spin polarization is gradually reversed. Hence the band inversion has a direct consequence on the nontrivial evolution of the
pseudo-spin polarization that is characterized by the number $c$. More specifically, the occurrence of band inversion leads to the
occurrence of a topological phase transition with $c_{S/S'}$ changing from 0 to $\pm 1$.

\begin{figure}
\begin{center}
  \includegraphics[width=0.42\textwidth]{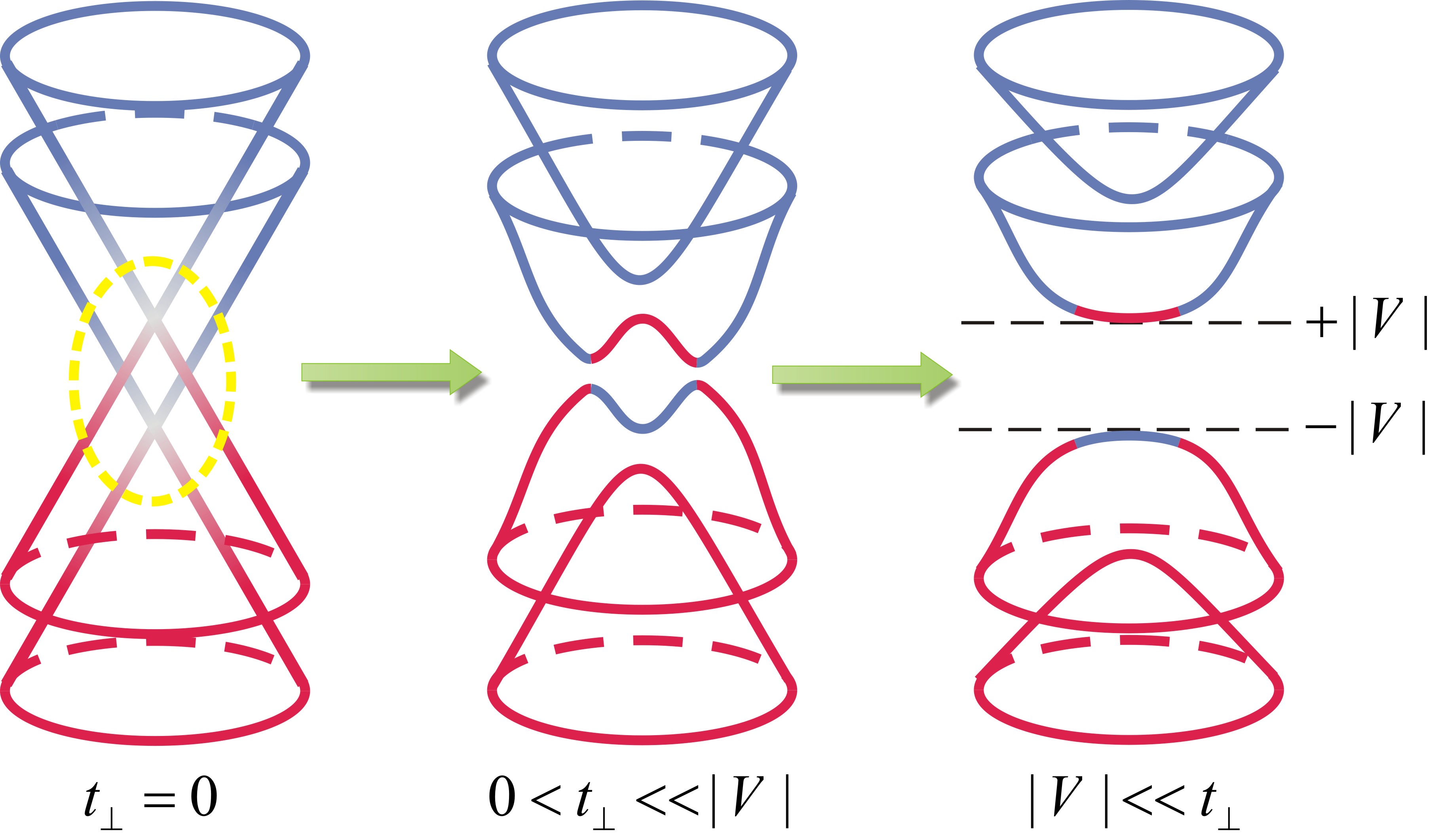}
\end{center}
\caption{Evolution of the band structure corresponding to the Hamiltonian \eqref{eq:K4} by tuning $\tp$, with fixed nonzero $V$. This
evolution illustrates the unusual aspect of the gap opening in BLG (see text for details).} \label{fig:inv_blg}
\end{figure}

In gapped BLG, a similar occurrence of band inversion can be best seen at the level of the single-valley four-component Hamiltonian
$H_{K}^{(4\times4)}$ --corresponding to two layers each with two sublattices-- from which the two-component one in Eq. \eqref{eq:ham_blgK}
is derived \cite{mccann_landau-level_2006}. Again, the symmetry makes it sufficient to consider one of the two valleys. Explicitly, we have
\begin{align}\label{eq:K4}
&H_{K}^{(4\times4)}(\bk) = \left(\begin{array}{cccc} V & \hbar v_F k_- & & \\ \hbar v_F k_+ & V & \tp & \\ & \tp & -V & \hbar v_F k_- \\ & &
\hbar v_F k_+ & -V \\
\end{array}\right),
\end{align}
where $\pm V$ are the electrostatic potential on two graphene layers, $\tp$ is the interlayer nearest-neighbor hopping energy (assuming
Bernal stacking of BLG), $k_{\pm} = k_x \pm ik_y$, $v_F = 3ta_0/2\hbar$ with $a_0$ and $t$ being the intralayer nearest-neighbor distance
and hopping energy. Note that, unlike Eq. \eqref{eq:ham_blgK}, this Hamiltonian is not dimensionless, and the parameters here are related to
those in Eq. \eqref{eq:ham_blgK} by defining $\Delta=V/\tp$ and redefining $\ell_0 \bk \rightarrow \bk$ with $\ell_0 = \hbar v_F / \tp$.

We imagine first that the coupling between the two graphene layers is turned off ($\tp=0$) such that, due to different electrostatic
potentials, the electronic band structure at one single valley consists of two shifted Dirac cones as illustrated in the left panel of Fig.
\ref{fig:inv_blg}. For clarity we mark with two different colors the initial conduction bands (blue) and valence bands (red) in Fig.
\ref{fig:inv_blg}. Then we gradually turn on $\tp$. Analogously to the turning on of $a$ in the BHZ model, a finite $\tp$ opens a gap in the
bulk and turns the system into an insulator (Fig. \ref{fig:inv_blg}). As a result, at low $k$, a small part of the valence band of one layer
is now ``glued" with the conduction band of the other layer to compose the new low-energy conduction band. The opposite happens to the new
low-energy valence band (see the middle panel of Fig. \ref{fig:inv_blg}). This is seemingly analogous to the band inversion we have seen in
the previous example. Upon increasing the magnitude of the interlayer coupling, the two high-energy bands are lifted further apart (the
splitting is roughly $2\tp$ when $\tp$ is large); the two low-energy bands each approach a parabolic shape, in agreement with Eq.
\ref{eq:ham_blgK}, while the inverted-band character nevertheless remains (see the right panel of Fig. \ref{fig:inv_blg}).

To summarize, the band inversion in gapped BLG, as
illustrated above, happens as long as $V\ne0$, and results
in a finite $c_{K/K'}$ that changes sign $\pm 1$ when $V$
changes sign. This is similar to the topological phase
transition exhibited by the BHZ model, but not identical,
since in the BHZ model $c_{S/S'}$ changes from 0 to 1 as
parameters are varied, and not from $-1$ to $+1$.

\section{Limitations of the analogy}\label{sec:lim}

One fundamental limitation of the analogy between graphene and topological insulators, with valley playing the role of spin, is that
valleys, which correspond to only parts of the Brillouin zone, are only defined in the energy window much smaller than the electron
bandwidth (fraction of eV). That is in contrast to electron spin which is well-defined throughout the Brillouin zone. In most cases, due to
the symmetry, the two valleys contribute oppositely to the topological property of the entire Brillouin zone -- Eq. \eqref{eq:cnum_blg} for
gapped BLG is an example. This results in overall topologically trivial graphene systems (according to the definition of a topological
insulator; in the present context, this means that the integral in Eq. (7) needs to be calculated over the entire Brillouin zone and not
only around one valley). This is certainly important since gapless edge states, if they exist at all, are not protected against intervalley
scattering. More importantly, even in the complete absence of intervalley scattering (in the bulk and at the edges) crucial differences
exist between the two systems at the level of single-valley models. These differences are the subject of this section.

\subsection{Characterization of bulk properties}\label{ssec:topo}

In the previous section we have employed Eq. \eqref{eq:cnum} for the two models that we are considering, suggesting that the numbers $c$
characterize the topological properties associated with bulk bands. We now go back to examine the validity of such characterizations, based
on a geometric interpretation of Eq. \eqref{eq:cnum}.

To begin with, we parametrize the general two-band Hamiltonian by setting $\hat{\bg} = (\sin\theta\cos\varphi, \sin\theta\sin\varphi,
\cos\theta)$, where $\theta$ and $\varphi$ are the angles defining the direction of $\bk$. It is straightforward to show that the integrand
of Eq. \eqref{eq:cnum} is equal to $\sin\theta (\dkx\theta\dky\varphi - \dkx\varphi\dky\theta)$. Recognizing that the factor in the
parenthesis is but the Jacobian $|\partial(\theta,\varphi)/\partial(k_x,k_y)|$, Eq. \eqref{eq:cnum} can be rewritten as
\begin{align}\label{eq:angle}
c = \frac{1}{4\pi}\iint \sin\theta d\theta d\varphi.
\end{align}
It is immediately clear that geometrically $c$ measures the solid angle (divided by $4\pi$) covered by $\hat{\bg}$ when $\bk$ runs over the
domain of interest.

Keeping this in mind, we can easily understand, by referring
back to Fig. \ref{fig:spin_all}, that $c = \pm 1$ for the
BHZ topological insulator because $\hat{\bg}$ covers the
full sphere once, while $c = \pm 1$ for the gapped BLG
because $\hat{\bg}$ covers a hemisphere (the upper or lower
hemisphere, depening on the sign of $\Delta$) exactly twice.
This fact reflects the crucial difference of how the vectors
$\hat{\bg}$ behave at $k \rightarrow \infty$ in the two
models. In the BHZ topological insulator case, $\hat{\bg}(k
\rightarrow \infty)$ converges identically to the south pole
regardless of the direction along which infinity is
approached, which allows for the compactification of the
infinite $k$-plane as a Riemann sphere. In this case, a
topologically nontrivial mapping from the compactified
$k$-plane to the sphere subtended by $\hat{\bg}$ is well
defined and $c$ is the topological invariant associated with
this mapping. In the gapped BLG case, however, we have a
different behavior: $\hat{\bg}(k \rightarrow \infty)$ sits
on the equator and varies with the polar angle of $\bk$,
preventing a proper compactification of the $k$-plane.
Consequently the numbers $c$ obtained , albeit nonzero,
cannot be identified as topological invariants \cite{note2} as in the previous case.

Despite this important difference, an alternative
interpretation of Eq. \eqref{eq:cnum} in terms of Berry
phase and Hall conductivity remains valid, as we proceed to
discuss. This indicates that the number $c$ has a
well-defined physical (albeit not topological) meaning
associated with the properties of the bulk. The Berry phase
acquired by an electron transported adiabatically along a
closed trajectory $\mathcal{C}$ in the momentum space is
given by
\begin{align}\label{eq:Bphase}
\gamma = i\oint_\mathcal{C} \langle u(\bk) | \nabla_k u(\bk) \rangle \cdot d\bk\;,
\end{align}
where $u(\bk)$ is the periodic part of the Bloch wave function and can be obtained by taking the eigenstate of the Bloch Hamiltonian
$H(\bk)$. Using Stokes' theorem it is easy to shown that the above formula is equivalent to
\begin{align}\label{eq:Bphase2}
\gamma = \frac{1}{2}\int_{\mathcal{S}_\mathcal{C}} d^2 k\; 4\, \imag\left( \langle \dky u(\bk) | \dkx u(\bk) \rangle \right),
\end{align}
with $\mathcal{S}_\mathcal{C}$ the area enclosed by $\mathcal{C}$. For $H(\bk) = \bg(\bk)\cdot\bsigma$ with $\hat{\bg}(\bk)$ =
($\sin\theta\cos\varphi$, $\sin\theta\sin\varphi$, $\cos\theta$) (the same parametrization as used above; we consider a fully gapped system
where $|\bg(\bk)|\ne0$ for all $\bk$), the lower band eigenstate can be written as $u(\bk) = (\sin\frac{\theta}{2}e^{-i\varphi},
-\cos\frac{\theta}{2})^T$. Then it is straightforward to show that the integrand in Eq. \eqref{eq:Bphase2} is again equal to $\sin\theta\,
|\partial(\theta,\varphi)/\partial(k_x,k_y)|$, hence
\begin{align}\label{eq:Bphase_cnum}
\gamma = 2\pi c\;.
\end{align}
It follows from Eq. \eqref{eq:Bphase2} that $\gamma$ can
also be identified as the Hall conductivity (in units of
$e^2/2\pi h$) contributed by all the occupied Bloch states
\cite{thouless_quantized_1982}. Therefore, $c_{K/K'}$ in Eq.
\eqref{eq:cnum_blg}, despite not representing a true
topological invariant, still corresponds to the
valley-specific Hall conductivity
\cite{semenoff_condensed-matter_1984,
xiao_valley-contrasting_2007, zhang_spontaneous_2011}, (in
units of $e^2/h$) when the Fermi energy lies in the bulk
band gap (i.e., $c$ is the contribution of the occupied
states in a given valley to the bulk Hall conductivity).
Similarly, of course, $c_{S/S'}$ in the BHZ model represent
the spin-specific Hall conductivity, next to being
well-defined topological invariants.

\subsection{Bulk-edge correspondence}

For the BHZ topological insulators $c_{S/S'}$ are topological invariants. According to the principle of bulk-edge correspondence, this
implies the presence of one pair of spin-helical gapless edge modes at each edge \cite{bernevig_quantum_2006, zhou_finite_2008} -- their
helicity follows from the opposite signs of $c_{S}$ and $c_{S'}$. In contrast, a similar conclusion cannot be drawn for the valley-specific
$c_{K/K'}$, since these quantities are not topological invariants. In general, therefore, we cannot expect bulk-edge correspondence in this
case. In the particular example of gapped BLG, this leads, as we will show, to the possibility that the number of gapless edge modes is
dependent on boundary conditions.

\begin{figure}
\begin{center}
  \includegraphics[width=0.45\textwidth]{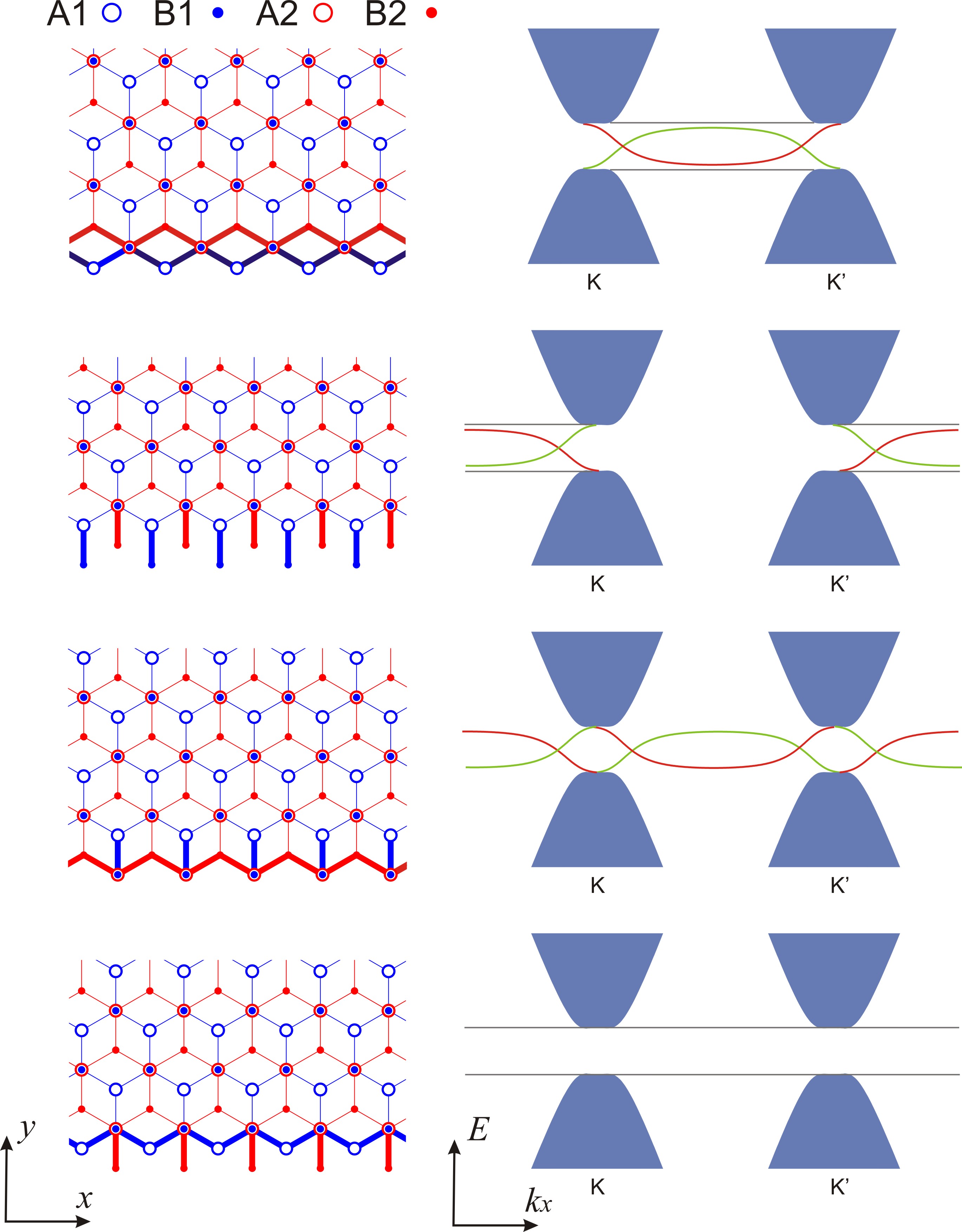}
\end{center}
\caption{Examples of defect-free edges (left panels) of BLG
and their corresponding electronic spectra (right panels;
bulk states in blue, subgap states at opposite edges in red
and green, dispersionless edge states in gray). None of the
edges shown here couples valleys, which ensures the
single-valley models being valid and definite. The numbers
of the subgap edge modes nevertheless are different for
different edge configurations, confirming the absence of
bulk-edge correspondence for individual valleys. Note that
in the right panels, (dispersion relation) lines extending
to the left and to the right actually connect at the
Brillouin zone boundary.} \label{fig:edges}
\end{figure}

To analyze the presence of edge states in gapped BLG we solve the wave equation for different edge structures (that do not couple states in
different valleys), to which corresponding appropriate boundary conditions are associated. Possible boundary conditions for the
single-valley models can be derived from the constraint of vanishing probability current across the boundary (we comment on their physical
realizations shortly later). Again we use one single valley Hamiltonian, described by Eq. \eqref{eq:ham_blgK}, as our example. Let us write
the two-component wavefunction as $\Psi(\bm{r})=(\psi_1(\bm{r}),\psi_2(\bm{r}))^T$. From the single valley continuity equation
$\frac{\partial}{\partial t}(\Psi^\dagger\Psi)+\nabla\cdot\bm{j}=0$, we obtain the probability current density
\begin{align}\label{eq:j}
\bm{j}(\bm{r}) = [i(\psi_1^*\partial_-\psi_2 + \psi_2^*\partial_+\psi_1) + c.c.]\hat{x}& \nonumber\\ + [(\psi_1^*\partial_-\psi_2 -
\psi_2^*\partial_+\psi_1) + c.c.]\hat{y}&\;,
\end{align}
where $\partial_\pm = \partial_x \pm i\partial_y$ and $c.c.$ stands for complex conjugation. Without losing generality, let us suppose we
have a semi-infinite sample occupying the upper half plane with its edge along $y=0$. The requirement that no current should flow through
the boundary implies
\begin{align}\label{eq:jy}
j_y(y=0) = [(\psi_1^*\partial_-\psi_2 - \psi_2^*\partial_+\psi_1) + c.c.]|_{y=0} = 0\;.
\end{align}
This equation can be satisfied in various settings, each accounting for one possible boundary condition. For example, we can easily see that
any of the following conditions satisfy the above equation,
\begin{align}
\psi_1|_{y=0} = (\partial_+\psi_1)|_{y=0} &= 0\;; \label{eq:BC1}\\ \psi_2|_{y=0} = (\partial_-\psi_2)|_{y=0} &= 0\;; \label{eq:BC2}\\
\psi_1|_{y=0} = \psi_2|_{y=0} &= 0\;; \label{eq:BC3}\\ (\partial_+\psi_1)|_{y=0} = (\partial_-\psi_2)|_{y=0} &= 0\;. \label{eq:BC4}
\end{align}

In principle, the boundary conditions obtained this way should be verified physically by finding corresponding terminations of the BLG
lattice, which is in general a formidable job. As a matter of fact, the examples given in Eqs. (\ref{eq:BC1}-\ref{eq:BC4}) can be derived
alternatively from the tight-binding models by considering specific sublattices that are missing at a sample edge. This latter approach
follows in spirit Brey and Fertig \cite{brey_electronic_2006} who dealt with single-layer graphene nanoribbons, and has been discussed in
details for the case of BLG in Ref. \cite{li_marginality_2010}. At this point, without entering details, we simply list the physical
realizations of the boundary conditions (\ref{eq:BC1}-\ref{eq:BC4}) in the same order in the left panels of Fig. \ref{fig:edges}. Note in
particular that all the edges shown are parallel to a zigzag configuration, which leaves the valleys uncoupled and hence allows the
applications of the single-valley models.

Also illustrated in Fig. \ref{fig:edges} are the bulk band profiles (blue areas) and the dispersion relations of edge states (red and green
curves for two opposite edges) corresponding to the specific type of edges shown in the same row of the figure. These spectra of electronic
states can be calculated either from solving the single-valley models with boundary conditions discussed above, or from exact
diagonalization of tight-binding models including edges -- the results of the two approaches agree quantitatively
\cite{li_marginality_2010}. One can immediately see that, as a most important fact, the number of subgap edge modes per valley per edge
explicitly depends on the edge configurations, and can be $1$, $2$ or $0$. Therefore, a naive extension of bulk-edge correspondence to the
individual valleys, which would predict the above number to be $1$ independent of the specific boundary, does not hold for gapped BLG.

We close this section with two remarks. First, our focus in
this paper centers on the subgap edge states, which are
dispersive and may contribute significantly to the subgap
conductance. Nevertheless, there can be another set of edge
states (also depending on boundary conditions) that are
dispersionless and form flat bands
\cite{castro_localized_2008} (shown as gray lines in Fig.
\ref{fig:edges}). Indeed, the presence of both types of edge
states as a whole has a well-defined topological origin from
the bulk \cite{ryu_topological_2002}, and can be formulated
in terms of the Zak's geometric phase associated with the
Bloch states for which an explicit account of the
crystallographic periodicity of the BLG lattice is necessary
\cite{Delplace_zak_2011}. Second, it is an interesting
problem to understand the absence of valley-specific
bulk-edge correspondence in the context of Laughlin's
gedanken experiment -- the inequality (for each valley) of
the quantized Hall conductivity and the number of gapless
edge modes seems to be a paradox. The key to resolve this
puzzle is to notice that the valley quantum number, unlike
charge, is not conserved during the adiabatic insertion of
flux which is normally considered in a Laughlin's gedanken
experiment. Such non-conservation of the valley quantum
number prevents a simple relation between the quantized
valley-specific Hall conductivity and the number of
valley-polarized gapless edge modes
\cite{li_marginality_2010, Martin_notes_2010}.

\section{Marginal topological character of graphene systems}

Although the analogy between valley-based graphene systems
and spin-based topological insulators has significant
limitations, the bulk topological properties of graphene
systems, as we show below, play nevertheless an important
role, with experimental consequences. To emphasize the
connection of graphene and topological insulators, and at
the same time to stress that there are important limitations
to this connection, we refer to graphene as to a ``marginal
topological insulator". We borrow the term ``marginal" from
Volovik to characterize the occurrence of noncompact
momentum space in the case of massive Dirac fermions
\cite{volovik_universe_2003}. In the following section we
illustrate that such a marginal character has direct
experimental implications.

\subsection{Domain wall}

The noncompact momentum space for single-valley models can
be sometimes ``cured" to obtain well-defined nontrivial
topological properties. One such case is a smooth
topological defect, such as a domain wall
\cite{martin_topological_2008, semenoff_domain_2008,
teo_topological_2010, jung_valley-hall_2011}.

\begin{figure}
\begin{center}
  \includegraphics[width=0.37\textwidth]{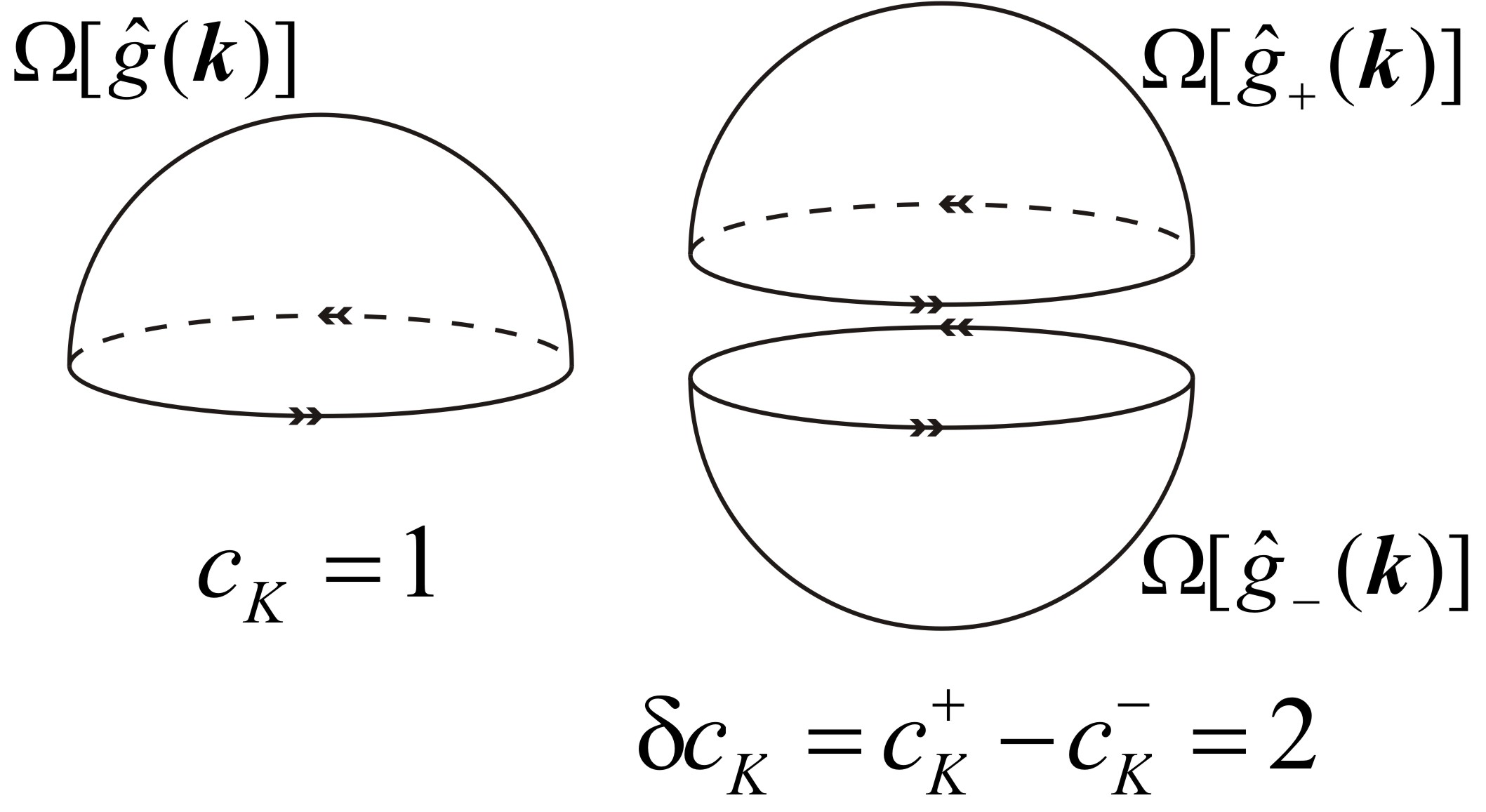}
\end{center}
\caption{Illustration of the mapping $\hat{\bg}(\bk)$ for gapped BLG. The left panel shows the solid angle $\Omega$ that $\hat{\bg}$ covers
when $\bk$ runs over the whole plane, the double arrow signifying the double covering of the (upper) hemisphere. The right panel shows how
two marginal topological mappings can be ``glued" to form a well-defined topological mapping, which is the case for a BLG-BLG domain wall
with opposite gap ($\Delta$) signs on the two sides (the sub-/super-scripts $+$ and $-$ denote the domains). $c_K$ in the left panel case is
not a topological invariant, but $\delta c_K$ in the right panel case is.} \label{fig:mappings}
\end{figure}

In the case of gapped BLG, a domain wall can be realized controllably by gating two adjacent areas of a sample with opposite polarities
(i.e. opposite signs of $\Delta$) \cite{martin_topological_2008}. The resulting nontrivial topological character that can be ascribed to an
individual valley can be understood as follows. We have discussed in the previous section that the $\hat{\bg}$ vectors --corresponding to
the pseudo-spin polarization of Bloch states-- cover a hemisphere twice when $\bk$ runs over the 2D $k$-plane in the case of a single valley
of uniformly-gapped BLG (see the left panel of Fig. \ref{fig:mappings}). It is also clear that, for a specific valley, the hemisphere which
$\hat{\bg}$ covers is determined by the sign of $\Delta$. It follows that for a domain wall across which $\Delta$ changes sign, the bulk
$\hat\bg$ vectors of the two domains cover two opposite hemispheres (each twice) which seamlessly connect on the equator at $k\rightarrow
\infty$ (where $\Delta$ in the Hamiltonians becomes irrelevant; see the right panel of Fig. \ref{fig:mappings}). In this way the momentum
space --within the same valley-- is compactified as a 2D sphere and the topological invariant characterizing the mapping from such a sphere
to the parametric $\hat{\bg}$-sphere is evidently nontrivial. In fact, this topological invariant is given by $\delta c_{\tau} =
c_{\tau}^+-c_{\tau}^-$ ($\tau=K,K'$), which is the difference of the number $c_{\tau}$'s for the two domains.

For a BLG domain wall, therefore, $\delta c_{K/K'} = \pm 2$,
and the difference {\em is} a topological invariant.
Moreover, it is this difference that defines the number of
the valley-polarized chiral modes via the spectral asymmetry
theorem \cite{volovik_universe_2003}. While the theorem
readily applies to smooth -- quasiclassical -- interfaces
such as domain walls, its conditions are in general violated
at the sharp interface between graphene and vacuum, even
when the interface does not mix valleys. The problem is that
one cannot adiabatically interpolate between the spectrum of
gapped BLG to the spectrum of vacuum, while one easily can
in the case of an interface between the regions with $\Delta
>0$ and $\Delta < 0$. Indeed this is what has been
analytically and numerically verified by Martin, Blanter and
Morpurgo \cite{martin_topological_2008}. Moreover, the
presence of these ``interface" gapless modes --unlike those
discussed previously for plain BLG edges-- does not depend
on microscopic details such as the crystallographic
configuration of the domain wall, which confirms the
topological origin of these modes.

\subsection{Gapped BLG with Disordered Edges}
\label{sec:disorder}

The marginality of single-valley models implies, on the one hand, that a small perturbation in the Hamiltonian or boundary conditions can
have a substantial effect on presence or absence of the gapless modes (e.g. see Ref. \cite{yao_edge_2009}) -- this is merely a restatement
of the absence of bulk-edge correspondence for individual valleys. On the other hand, considering a realistic graphene device where the
edges of the sample are inevitably under the influence of disorder, physical properties will necessarily be subject to ensemble averages,
such that, at sufficiently strong disorder, universal behaviors are expected to occur. It is thus interesting to ask in this context
whether, or how, the marginal bulk property can be manifested in the subgap regime.

\begin{figure}
\begin{center}
  \includegraphics[width=0.4\textwidth]{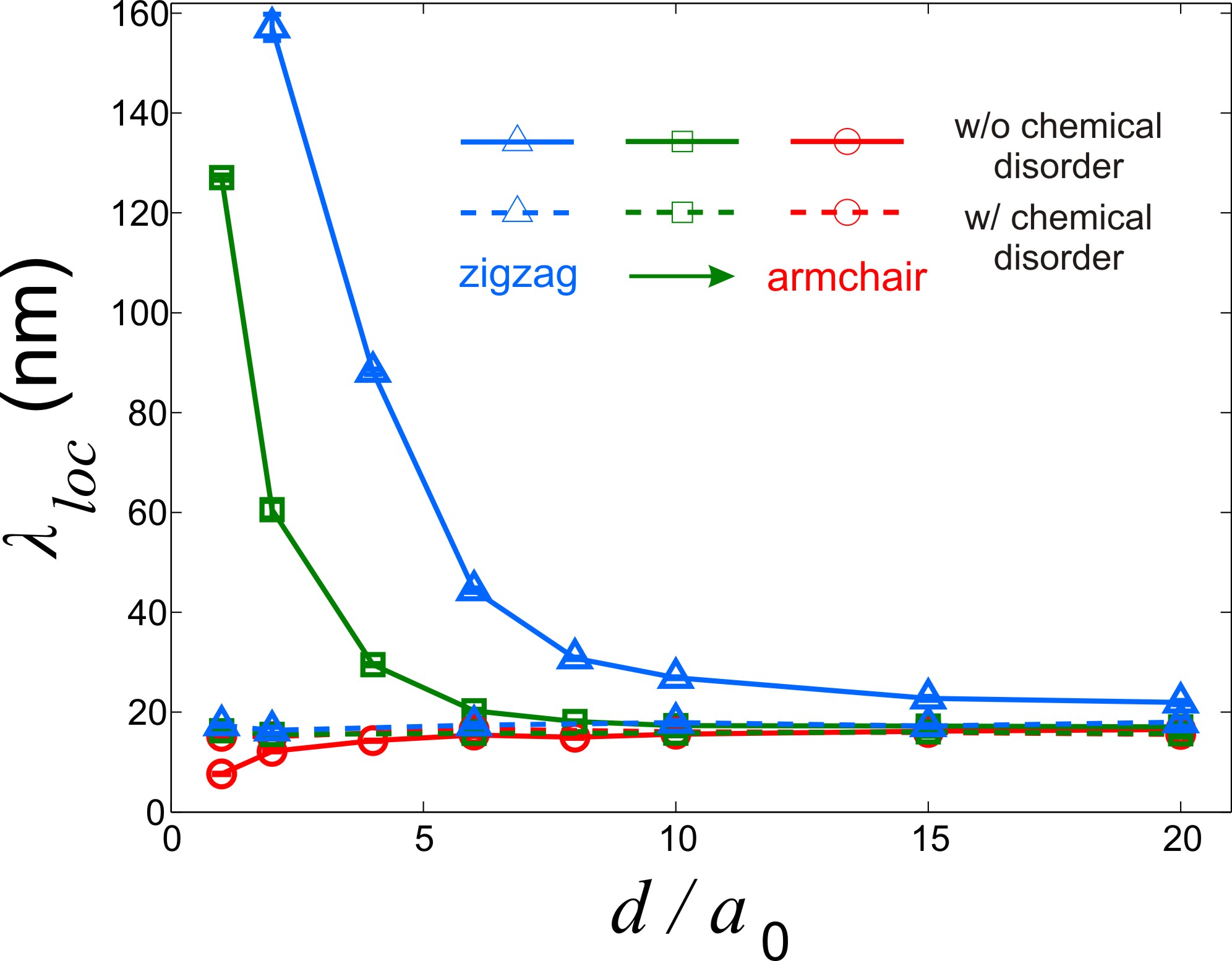}
\end{center}
\caption{The localization length $\lambda_{loc}$ of
low-energy edge states as a function of edge roughness depth
$d$ (divided by $a_0$, the intralayer nearest-neighbor
distance of graphene). The simulations to extract
$\lambda_{loc}$ are carried out starting from different
ordered edges, with or without chemical disorder. In all
cases, sufficiently strong disorder leads to the same value
of $\lambda_{loc}$ regardless of any detail. This universal
localization length, $\lambda^*$, is around 20 nm for the
current case shown here, corresponding to a gap of
approximately 100 meV.} \label{fig:lls}
\end{figure}

Such a question has been addressed by Li et al.
\cite{li_topological_2011} with the motivation to
investigate the ``extra" contribution by the edge states to
the subgap conduction in realistic experiments on gapped
BLG. It is found that indeed strong edge disorder leads to
universal presence of subgap edge states which are localized
with fairly long localization length. To be specific, Fig.
\ref{fig:lls} shows how the localization length
$\lambda_{loc}$ --extracted numerically with the aid of the
scattering theory \cite{buttiker_symmetry_1988,
tworzydlo_sub-poissonian_2006}-- depends on the strength of
edge disorder ($d$ denotes edge roughness depth; solid lines
represent the effect of structural disorder only and broken
lines represent the effect of structural plus chemical
disorder; see Ref. \cite{li_topological_2011} for details),
with various choices of initially ordered (e.g. zigzag,
armchair) edges, and a fixed gap size at an experimentally
accessible value. For the cases where the starting edges
accommodate pre-existing edge states (counter-propagating
for two valleys), gradually turning on disorder leads to
localization and a shrinking in the transverse size of the
edge states (see the solid lines in blue and green) -- this
is certainly expected because of the lack of protection for
the edge states from back-scattering in the system. What is
unexpected is that when starting from initial edges where no
edge state exists even in the absence of disorder, disorder
starts to introduce (localized) edge states inside the gap.
Moreover at relatively weak disorder, increasing disorder
enhances their $\lambda_{loc}$ (see the solid line in red).
Most interestingly, the localization length in all cases
converges to the same value at sufficiently strong disorder
(see the solid lines in the large-$d$ limit, and also the
broken lines where dominant chemical disorder leads to very
fast convergence even at small $d$). The universal value
$\lambda^*$ depends only on the gap size ($\lambda^* \propto
1/\sqrt{\Delta}$; it is around $20$ nm in the cases
illustrated in Fig. \ref{fig:lls} corresponding to a gap of
approximately 100 meV). These values of the localization
length can have experimental consequences in real devices
\cite{li_topological_2011}.

\begin{figure}
\begin{center}
  \includegraphics[width=0.45\textwidth]{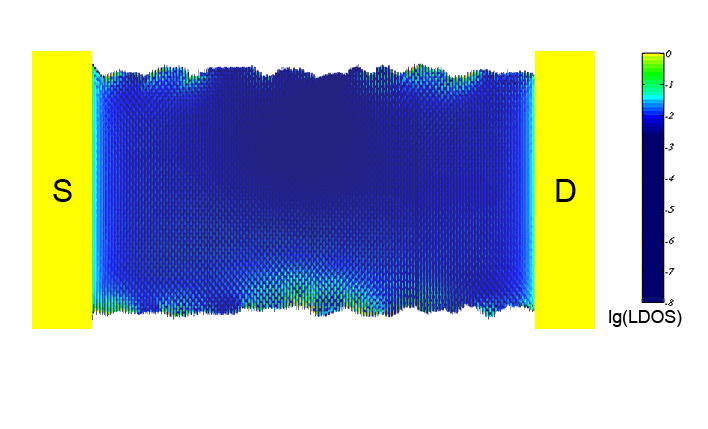}
\end{center}
\caption{Local density of states at zero energy (center of the gap) in a typical two terminal device of gapped BLG with disordered edges.
Edge states are universally present in the subgap regime with long localization length.} \label{fig:ldos}
\end{figure}

Thus subgap edge states with a long localization length are
generically present at disordered edges of gapped BLG (see
Fig. \ref{fig:ldos}). These states can manifest themselves
in transport experiments at finite temperature. In
particular, when the subgap transport is dominated by
one-dimensional variable range hopping along the edges
(which is the case if the hopping through the bulk disorder
are sufficiently suppressed), the temperature dependence of
the conductance is of the form $G \propto
\exp[-(T^*/T)^{1/2}]$, with $T^*$ the characteristic energy
scale that determines the low temperature transport
behavior. For practically relevant parameters, $T^*$ is
proportional to, but approximately one order of magnitude
smaller than the size of the bulk gap
\cite{li_topological_2011}. Indeed in very clean gapped BLG
devices, in which the bilayer is suspended and not in
contact with a substrate, an excess conductance, thermally
activated with a characteristic energy one order of
magnitude smaller than the gap, has been observed
experimentally \cite{weitz_broken-symmetry_2010}. These
observations are in agreement with our predictions.

To verify the link between presence of the weakly localized subgap edge states and the marginal bulk properties, we performed a comparative
study between the current BLG system and a half-filled square lattice with different on-site energies on nearest neighbors. Such a square
lattice is fully-gapped but has a topologically trivial band structure \cite{li_topological_2011}. It turns out that in the latter system no
subgap edge states are present neither in the ordered nor in the disordered limit; subgap transport is solely due to weak direct tunneling
between contacts which is even further suppressed in the presence of edge disorder. The sharp contrast between the two systems strongly
suggests that the universal features in terms of the (localized) subgap edge states in gapped BLG is indeed a manifestation of its
non-trivial marginal topological bulk properties.

\section{Conclusion}
To conclude, we have carefully explored the apparent analogy
between gapped graphene system and topological insulators,
with the central element of the correspondence being the
valley degree of freedom in the former and electron spin in
the latter. We find that the strong formal resemblance is
fundamentally limited by the different topological
structures in the electron spectrum underlying the two
systems. We have used gapped BLG as a concrete example to
facilitate the comparison, and more importantly, to
demonstrate how the ``marginal" topological properties
associated with valleys can manifest themselves
significantly in experiments. Yet, such marginal
characteristics are generic for graphene systems, and can be
readily seen by considering other graphene multi- and single
layers.

\section{Acknowledgement}
This work has been supported by the Swiss National Science Foundation
Projects No. 200020-121807 and No. 200021-121569, by the Swiss Centers
of Excellence MaNEP and QSIT, and the European Network NanoCTM. The work
of I.M. was carried out under the auspices of the National Nuclear
Security Administration of the U.S. Department of Energy at Los Alamos
National Laboratory under Contract No. DE-AC52-06NA25396 and supported
by the LANL/LDRD Program.


%

\end{document}